\documentclass[a4,12pt]{article}
\usepackage[dvips]{graphicx,psfrag}
\usepackage{float}
\usepackage{amsmath,amsthm,amssymb}

\def\ls{\lower0.5ex\hbox{$\buildrel >\over{\scriptstyle\sim}$}} 
\def\rs{\lower0.5ex\hbox{$\buildrel <\over{\scriptstyle\sim}$}} 
\allowdisplaybreaks
\begin{document}
\pagestyle{empty} \setlength{\footskip}{2.0cm}
\setlength{\oddsidemargin}{0.5cm}
\setlength{\evensidemargin}{0.5cm}
\renewcommand{\thepage}{-- \arabic{page} --}
\def\mib#1{\mbox{\boldmath $#1$}}
\def\bra#1{\langle #1 |}  \def\ket#1{|#1\rangle}
\def\vev#1{\langle #1\rangle} \def\dps{\displaystyle}
\newcommand{\fcal}{{\cal F}}
\newcommand{\gcal}{{\cal G}}
\newcommand{\ocal}{{\cal O}}
\newcommand{\El}{E_\ell}
\renewcommand{\thefootnote}{$\sharp$\arabic{footnote}}
\newcommand{\W}{{\scriptstyle W}}
 \newcommand{\I}{{\scriptscriptstyle I}}
 \newcommand{\J}{{\scriptscriptstyle J}}
 \newcommand{\K}{{\scriptscriptstyle K}}
%
 \def\thebibliography#1{\centerline{REFERENCES}
 \list{[\arabic{enumi}]}{\settowidth\labelwidth{[#1]}\leftmargin
 \labelwidth\advance\leftmargin\labelsep\usecounter{enumi}}
 \def\newblock{\hskip .11em plus .33em minus -.07em}\sloppy
 \clubpenalty4000\widowpenalty4000\sfcode`\.=1000\relax}\let
 \endthebibliography=\endlist
 \def\sec#1{\addtocounter{section}{1}\section*{\hspace*{-0.72cm}
 \normalsize\bf\arabic{section}.$\;$#1}\vspace*{-0.3cm}}
\def\secnon#1{\section*{\hspace*{-0.72cm}
 \normalsize\bf$\;$#1}\vspace*{-0.3cm}}
 \def\subsec#1{\addtocounter{subsection}{1}\subsection*{\hspace*{-0.4cm}
 \normalsize\bf\arabic{section}.\arabic{subsection}.$\;$#1}\vspace*{-0.3cm}}
\vspace*{-1.7cm}
\begin{flushright}
$\vcenter{
\hbox{{\footnotesize FUT and TOKUSHIMA Report}}
{ \hbox{(arXiv:1206.2413)}  }
}$
\end{flushright}

\vskip 1.4cm
\begin{center}
\hspace*{-0.6cm}
$\mbox{\large\bf Optimal-observable Analysis of Possible Non-standard}$

\vskip 0.18cm
{\large\bf Top-quark Couplings in $\mib{pp\to t \bar{t}X \to \ell^+ X'}$}

\end{center}

\vspace{0.9cm}
\begin{center}
\renewcommand{\thefootnote}{\alph{footnote})}
Zenr\=o HIOKI$^{\:1),\:}$\footnote{E-mail address:
\tt hioki@ias.tokushima-u.ac.jp}\ and\
Kazumasa OHKUMA$^{\:2),\:}$\footnote{E-mail address:
\tt ohkuma@fukui-ut.ac.jp}
\end{center}

\vspace*{0.4cm}
\centerline{\sl $1)$ Institute of Theoretical Physics,\
University of Tokushima}

\centerline{\sl Tokushima 770-8502, Japan}

\vskip 0.2cm
\centerline{\sl $2)$ Department of Information Science,\
Fukui University of Technology}
\centerline{\sl Fukui 910-8505, Japan}

\vspace*{2.3cm}
\centerline{ABSTRACT}

\vspace*{0.2cm}
\baselineskip=21pt plus 0.1pt minus 0.1pt
Possible non-standard top-quark interactions with the gluon and the $W$ boson
induced by 
$SU(3) \times SU(2) \times U(1)$ gauge-invariant dimension-6 effective operators
are studied for hadron-collider experiments. Current limits on top-gluon couplings
are presented by using the latest experimental data of $t\bar{t}$ productions
at the Tevatron and the Large Hadron Collider (LHC). The optimal-observable procedure
is applied to the charged-lepton distributions in $pp\to t\bar{t}X \to \ell^+X'$
($\ell=e$ or $\mu$) at the LHC in order to estimate the expected statistical
uncertainties in measurements of those non-standard top-gluon and top-$W$
couplings that contribute to this process in the leading order.

\vskip 1.5cm

\centerline{(to appear in Physics Letters B)}

\vfill
PACS:\ \ \ \ 12.38.Qk,\ \ \  12.60.-i,\ \ \  14.65.Ha

\setcounter{page}{0}
\newpage
\renewcommand{\thefootnote}{$\sharp$\arabic{footnote}}
\pagestyle{plain} \setcounter{footnote}{0}

\sec{Introduction}
The Large Hadron Collider (LHC) is currently one of the most powerful facilities for probing
possible new physics beyond the standard model (BSM)~\cite{LHC}. In collider experiments
such as the LHC, there could be two kinds of approaches to new-physics studies: One is a direct search
for new particles which do not belong to the standard model (SM). The other is an indirect
search, which tries to find certain deviations from the SM predictions in processes having
only well-known particles in their initial and final states.

In the former case,
any non-standard particles once being discovered, we would get a lot of information
about new models which contain those particles. However, it is not that easy to enable
such high-energy collisions which are enough energetic to create new heavy particles. On the other
hand, in the latter case, observable deviations might appear even at an energy far below
a typical new-physics scale since they could be induced by quantum effects of non-SM heavy particles.
Of course a large number of events are needed in order to establish a new-physics evidence
at a statistically-significant level, because such
signals would generally be quite small, considering the remarkable success of the SM
in various phenomena below $O(10^2)$-GeV scale.

Standard particles have been satisfactorily ``re"discovered today except for the Higgs boson
at the LHC, including top quarks. This is the first time for the top quark to be measured at
a facility other than the Tevatron although almost twenty years have passed since its discovery
there~\cite{topdisc, Abachi:1995iq}.
This situation, in which more precise tests of top-quark properties
can be continued, is quite meaningful for new-physics studies after the Tevatron is shutdown:  
The top quark is the heaviest particle of all standard particles, indicating that precise
measurements of this quark might be able to open windows for the BSM~\cite{Kamenik:2011wt}.

In various direct searches of new particles up to today, any plausible candidates have not been
discovered yet, which definitely requires indirect BSM searches. Considering this circumstance, 
we pay attention in this work to $t\bar{t}$ productions followed by $t \to \ell^+ X$ ($\ell=e$ or $\mu$)
at hadron colliders, and investigate relevant non-standard top-quark couplings
$t\bar{t}g$, $t\bar{t}gg$ and $tbW$ based on the effective Lagrangian
which is constructed with SM particles alone.
So far, a number of authors have studied this issue in similar frameworks to extract BSM signals
at hadron colliders~\cite{Atwood:1992vj}-\cite{Zhang:2010dr}.
Since many of them are however focusing mainly on CP-violating observables,
we study top-quark couplings which induce  CP-conserving as well as CP-violating interactions.
We first give updated constraints on the top-gluon couplings coming from
the latest data on the $t\bar{t}$ total cross sections at the Tevatron and the LHC. Then
we perform an optimal-observable analysis (OOA) to estimate the expected statistical uncertainties
in measuring the top-gluon and top-$W$ couplings that contribute to
$pp \to t\bar{t}X \to \ell^+ X'$ under a linear
approximation for current and future LHC experiments, which is our final goal here.
\sec{Framework}
The effective-Lagrangian approach, in which its low-energy form reproduces the SM,
is one of the most promising methods to 
describe new-physics phenomena when the energy of our experimental facility is not high enough
to produce new particles. Assuming any non-SM particles too heavy to appear as real ones, the 
effective Lagrangian is given as
\begin{equation}\label{eq:efflag}
{\cal L}_{\rm eff}={\cal L}_{\rm SM} 
+ \frac{1}{{\mit\Lambda}^2} \sum_i \left(\,  {C_i \cal O}_i + {\rm h.c.} \,\right),
\end{equation}
where ${\cal L}_{\rm SM}$ is the SM Lagrangian, ${\cal O}_i$ mean $SU(3)\times SU(2)\times U(1)$
gauge-invariant operators of mass-dimension 6
involving only the SM fields~\cite{Buchmuller:1985jz, Arzt:1994gp}
and their coefficients $C_i$ parameterize virtual effects of new particles
at an energy less than the assumed new-physics scale 
${\mit\Lambda}$.\footnote{Since dimension-5 operators violate the lepton-number   
    conservation, they are not treated hereafter. Therefore, we deal with dimension-6 
    operators, which give the largest contributions in relevant processes.}\
In this framework, all the form factors related to $C_i$ are dealt with as constant parameters,
without supposing any specific new-physics models.

All the dimension-6 operators were initially listed by Buchm\"{u}ller and Wyler~\cite{Buchmuller:1985jz}.
However, it was pointed out that some operators there are related with others through
equations of motion, which means that they are not independent of each other~\cite{Grzadkowski:2003tf}.
Then, the whole related operators were rearranged to
get rid of this redundancy in Refs.~\cite{AguilarSaavedra:2008zc,Grzadkowski:2010es}.

Following the notation of Ref.~\cite{AguilarSaavedra:2008zc},
the relevant effective Lagrangian which describes non-standard
interactions of the third generation for the parton-level process
$q \bar{q} /gg  \to t \bar{t} \to \ell^+ X$ is given in~\cite{HIOKI:2011xx} as
\begin{alignat}{1}\label{eq:efflag_3rd}
 {\cal L}_{\rm eff} &={\cal L}_{t\bar{t}g,gg}+{\cal L}_{tbW}:  \\
 &  {\cal L}_{t\bar{t}g,gg} = -\frac{1}{2} g_s \sum_a \Bigl[\,\bar{\psi}_t(x)\lambda^a \gamma^{\mu}
    \psi_t(x) G_\mu^a(x)
    \bigl. \nonumber\\
  &\phantom{====}-\bar{\psi}_t(x)\lambda^a\frac{\sigma^{\mu\nu}}{m_t}\bigl(d_V+id_A\gamma_5\bigr)
  \psi_t(x)G_{\mu\nu}^a(x)\,\Bigr], \\
 &  {\cal L}_{tbW}  = -\frac{1}{\sqrt{2}}g 
  \Bigl[\,\bar{\psi}_b(x)\gamma^\mu(f_1^L P_L + f_1^R P_R)\psi_t(x)W^-_\mu(x) \Bigr. \nonumber\\
  &\phantom{====}+\bar{\psi}_b(x)\frac{\sigma^{\mu\nu}}{M_W}(f_2^L P_L + f_2^R P_R)
   \psi_t(x)\partial_\mu W^-_\nu(x) \,\Bigr], 
\end{alignat}
where $g_s$ and $g$ are the $SU(3)$ and $SU(2)$ coupling constants, 
$P_{L/R}\equiv(1\mp\gamma_5)/2$,
$d_V, d_A$ and $f_{1,2}^{L,R}$ are form factors defined as
\begin{alignat}{2}\label{eq:dvdadef}
 d_V &\equiv \frac{\sqrt{2}v m_t}{g_s {\mit\Lambda}^2} {\rm Re}(C^{33}_{uG\phi}),
 & \quad  d_A&\equiv \frac{\sqrt{2}v m_t}{g_s {\mit\Lambda}^2} {\rm Im}(C^{33}_{uG\phi}), \nonumber\\
  f_1^L&\equiv V_{tb}+C^{(3,33)*}_{\phi q}\frac{v^2}{{\mit\Lambda}^2},  
  & \quad  f_1^R&\equiv C^{33*}_{\phi \phi}\frac{v^2}{2{\mit\Lambda}^2},  \\
  f_2^L&\equiv -\sqrt{2} C^{33*}_{dW}\frac{v^2}{{\mit \Lambda}^2},
  & \quad  f_2^R&\equiv -\sqrt{2} C^{33}_{uW}\frac{v^2}{{\mit\Lambda}^2}, \nonumber
\end{alignat}
with $v$ being the Higgs vacuum expectation value
and $V_{tb}$ being ($tb$) element of Kobayashi--Maskawa matrix.
 In particular, $d_V$ and $d_A$ are the
so-called chromo\-magnetic- and chromo\-electric-dipole moments, respectively.

In the following calculations, 
we use the above effective Lagrangian for top-quark interactions and the usual SM Lagrangian 
for the other interactions which are not affected by top quarks hereafter.

\sec{Current bounds on ${\mib d_V}$ and ${\mib d_A}$}
Experimental constraints on $d_V$ and $d_A$ were derived in our preceding
works~\cite{Hioki:2009hm,HIOKI:2011xx}. Since LHC data of top-pair productions have
been updated, however, we also re-derive new limits on $d_V$ and $d_A$
combining the recent data from the Tevatron for $\sqrt{s}=1.96$ TeV
and the LHC for $\sqrt{s}=7$ TeV assuming $m_t=172.5$ GeV:
\begin{alignat*}{2}
 \sigma_{\rm exp} &= 7.50\pm 0.48 ~{\rm [pb]},& \quad  &\text{({CDF}~\cite{CDFnote})}\\
  &= 7.56^{+0.63}_{-0.56}  ~{\rm [pb]}, & \quad  &\text{({D0}~\cite{D0:2011cq})}\\
  &= 187^{+22}_{-21}  ~{\rm [pb]},& \quad  &\text{({ATLAS}~\cite{Aad:2012qf})}\\
  &= 165.8 \pm {13.3}  ~{\rm [pb]}.& \quad  &\text{({CMS}~\cite{CMS})}
 \end{alignat*}
The corresponding theoretical predictions in the SM taking into account QCD 
higher-order 
corrections have been given by several authors~\cite{Cacciari:2003fi}--\cite{Kidonakis:2010dk}. 
Instead of averaging them, 
we take here the following values: 
\begin{alignat}{2} 
\sigma_{\rm SM} 
  &=7.29^{+0.79}_{-0.85}~{\rm [pb]} & \quad  &{\rm for\ Tevatron}, \label{sigSM1}
  \\ 
  &=165^{+11}_{-16}     ~{\rm [pb]} & \quad  &{\rm for\ LHC},  \label{sigSM2}
\end{alignat} 
which were used in ~\cite{CDFnote,Aad:2012qf}. They are consistent with the 
others, and
therefore this choice does not cause any serious problem
to our results shown below except for some tiny difference.

Adding those theoretical uncertainties to the above experimental errors 
quadratically, 
we compare thus-obtained ``effective'' experimental data 
\begin{alignat*}{2} 
 \sigma_{\rm exp}^{\rm eff} &=7.50^{+0.92}_{-0.98}~{\rm [pb]}& \quad  &{\rm for\ CDF,}\\
  &=  7.56^{+1.01}_{-1.02}~{\rm [pb]}& \quad  &{\rm for \ D0,}\\
  &=  187^{+25}_{-26}  ~{\rm [pb]}& \quad  &{\rm for \ ATLAS,}\\
  &=  166^{+17}_{-21} ~{\rm [pb]} & \quad  &{\rm for \ CMS,}
 \end{alignat*}
as our inputs with
\begin{equation}
\sigma = \sigma_{\rm SM} + {\mit\Delta}\sigma(d_V, d_A),
\end{equation}
where $\sigma_{\rm SM}$ represents the central values in the above equations (\ref{sigSM1}, \ref{sigSM2}),
${\mit\Delta}\sigma(d_V, d_A)$ is the non-SM contribution calculated based
on the effective Lagrangian, and the explicit forms of the corresponding parton-level quantities,
i.e., ${\mit\Delta}\sigma_{q\bar{q}}(d_V, d_A)$ and ${\mit\Delta}\sigma_{gg}(d_V, d_A)$ are
found in the first article of Ref.~\cite{Hioki:2009hm}.

\begin{figure}[!h]
\vspace*{-0.2cm}
\begin{center}
   \includegraphics[scale=1.0,clip]{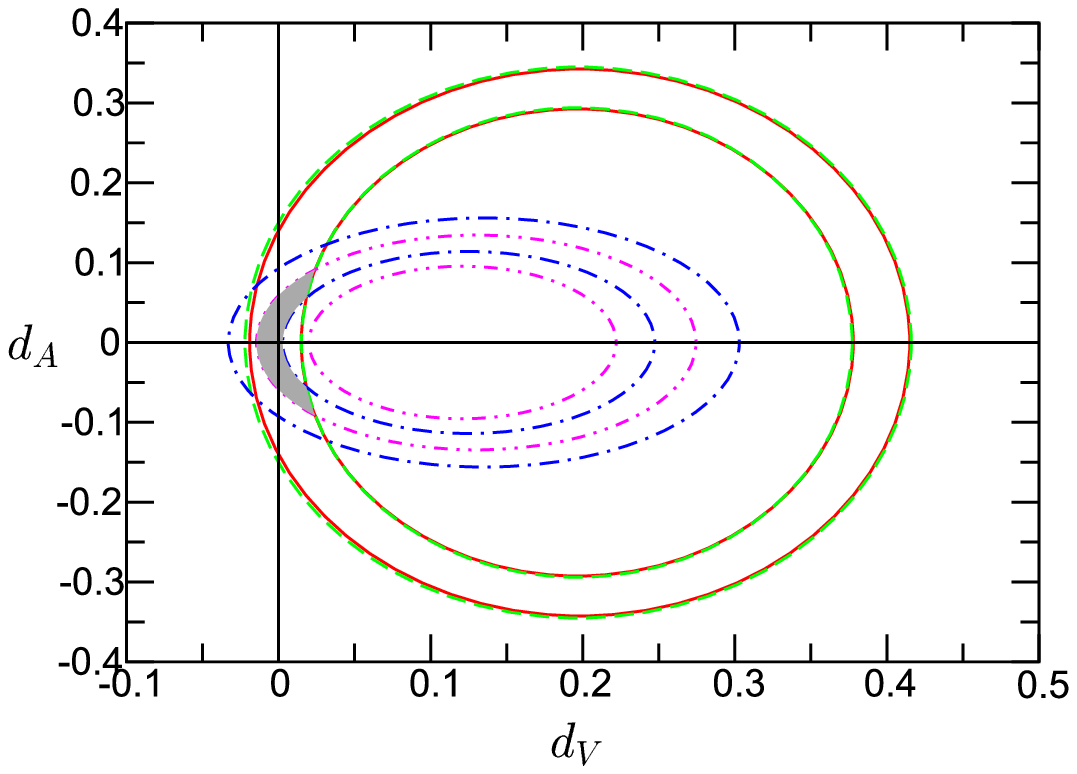}
\vspace*{-0.3cm}
\caption{The $d_{V,A}$ region allowed by the Tevatron and LHC data altogether (the shaded part).
The solid curves, the dashed curves, the dash-dotted curves and the dash-dot-dotted curves are
respectively from CDF, D0, ATLAS and CMS data.}\label{allowed}
\end{center}
\end{figure}

Numerical analyses are performed the same manner (including the parton-distribution functions) as
in \cite{Hioki:2009hm} and the whole result is shown in Fig.\ref{allowed}. The shaded part there
expresses the region of $d_{V}$ and $d_A$ allowed by the above Tevatron and LHC data altogether.
This shows clearly that both $d_V$ and $d_A$ are well restricted though there remains some small
room for non-SM contribution.

%
%

\sec{Optimal-observable analysis}
%
The optimal-observable analysis (OOA) is a method that could systematically estimate
the expected statistical uncertainties of measurable parameters. Here we do not go into
its detailed description and just show how to know the
uncertainties thereby, as more theoretical aspects of this 
method are presented in other papers~\cite{Atwood:1991ka}--\cite{Gunion:1996vv}. 

First, in order to apply the OOA to the process $pp\to t\bar{t} X \to \ell^+ X'$,
we express its differential cross section (the angular and energy distribution of
the charged-lepton $\ell^+$) which we derived in our previous paper~\cite{HIOKI:2011xx}
as follows:
\begin{alignat}{1}
\frac{d^2\sigma_{\ell}}{d E_{\ell} d \cos\theta_\ell} 
  &=  f_{\rm SM}(E_{\ell}, \cos\theta_\ell) 
  + d_V f_{d_V}(E_{\ell},\cos\theta_\ell) 
  + d_R f_{d_R}(E_{\ell},\cos\theta_\ell) \nonumber\\
  &\phantom{\equiv ~}+ d_V^2 f_{d_V^2}(E_{\ell}, \cos\theta_\ell)
  + d_A^2 f_{d_A^2}(E_{\ell}, \cos\theta_\ell)
  + \cdots,
\end{alignat}
where $f_{\rm SM}(E_{\ell}, \cos\theta_\ell)$ denotes the SM contribution,
all the other $f_I(E_{\ell}, \cos\theta_\ell)$ describe the non-SM terms
corresponding to their coefficients, and $d_R$ is defined as 
\begin{equation}\label{eq:drdef}
d_R\equiv {\rm Re}(f_2^R) M_W /m_t.
\end{equation}
The explicit forms of $f_I(E_{\ell}, \cos\theta_\ell)$ at the parton level are easily found in the
relevant formulas in~\cite{HIOKI:2011xx}.

Since the magnitude of $d_V$ and $d_A$ has been shown small as in Fig.\ref{allowed},
we neglect any contribution from terms quadratic (or higher) in those non-SM parameters
hereafter.\footnote{The quadratic terms of $f_{1,2}^{L,R}$ have already been
    neglected in our previous  paper~\cite{HIOKI:2011xx}. Indeed, studies at the Tevatron
    suggest that those contributions are tiny: See~\cite{Aaltonen:2012rz} for the latest data.}
Note that all $d_A$ contributions disappear under this linear approximation,
because there is no term proportional to $d_A$.
Thus, the angular and energy distributions of a decay lepton are written as
\begin{alignat}{1}
 \frac{d\sigma_{\ell}}{d \cos\theta_{\ell}} 
  &=  g_1(\cos\theta_\ell) + d_V ~g_2(\cos\theta_\ell),\\
 \frac{d\sigma_{\ell}}{d E_\ell} 
  &=  ~h_1(E_{\ell}) + d_V ~h_2(E_{\ell}) 
  + d_R~h_3(E_{\ell}),
\end{alignat} 
where $g_i(\cos\theta_\ell)$ and $h_i(E_{\ell})$ are given by
\begin{alignat*}{2}
  g_i(\cos\theta_\ell)&= \int d E_\ell~f_I(E_\ell,\cos\theta_\ell),  & \quad  
  h_i(E_{\ell})&=\int d \cos\theta_\ell~f_I(E_\ell,\cos\theta_\ell)
\end{alignat*}
with $i=1,2$ and 3 corresponding to $I={\rm SM}$, $d_V$ and $d_R$, respectively.
Here should be one comment about the angular distribution:
We can thereby probe exclusively the $d_V$ term, since any contribution from
$d_R$ disappears within our approximation as a result of the decoupling theorem found
in~\cite{Grzadkowski:1999iq}-\cite{Godbole:2006tq}.

Secondly, we calculate the following matrices:
\begin{alignat}{1}
 M_{ij}^c &\equiv \int d\cos\theta_{\ell} 
\frac{g_i(\cos\theta_\ell) g_j(\cos\theta_\ell)}
     {g_1(\cos\theta_\ell)}\ \ \ \ (i,\,j=1,\,2), \\
  M_{ij}^E&\equiv \int d E_\ell
  \frac{h_i(E_{\ell}) h_j(E_{\ell})}
     {h_1(E_{\ell})}\ \ \ \ (i,\,j=1,\,2,\,3)
\end{alignat}
and their inverse matrices $X_{ij}^{c,E}$, all of which are apparently symmetric.
Then the statistical uncertainties for the measurements of couplings
$d_V$ and $d_R$ could be estimated by
\begin{equation}\label{eq:uncertainty1}
| \delta d_V |= \sqrt{X_{22}^{c}\sigma_{\ell}/N_{\ell}}
=\sqrt{X_{22}^{c}/L}
\end{equation}
through the angular distribution, and
\begin{eqnarray}
&&| \delta d_V |= \sqrt{X_{22}^{E}\sigma_{\ell}/N_{\ell}}
=\sqrt{X_{22}^{E}/L}\,, \label{eq:uncertainty2}\\
&&| \delta d_R |= \sqrt{X_{33}^{E}\sigma_{\ell}/N_{\ell}}
=\sqrt{X_{33}^{E}/L}\label{eq:uncertainty3}
\end{eqnarray}
via the energy distribution,
where $\sigma_{\ell}$, $N_{\ell}$ and $L$ denote
the total cross section, the number of events and the integrated luminosity
for the process, respectively.

We are now ready to carry out necessary numerical calculations.
Below we show the elements of $M^{c,E}$ computed for $\sqrt{s}=7,8,10$, and 14 TeV, assuming
$m_t=173$ GeV.\footnote{We performed our analysis
    of the $t\bar{t}$ total cross section for $m_t=172.5$ in \S 3 since this value was
    assumed in the Tevatron and LHC data. Hereafter, however, we would like to use
    $m_t=173$ GeV considering the latest world average \cite{Brandt:2012ui}.}
\begin{description}
 \item[(1)] The angular distribution 
 \begin{description} 
 \item[(1-1)] $\sqrt{s}=7$ TeV 
 \begin{alignat*}{3}
  M^c_{11}&= 23.102, & \quad  M^c_{12}&= -245.412,  
  & \quad  M^c_{22}&=2607.340.
\end{alignat*}
 \item[(1-2)] $\sqrt{s}=8$ TeV
  \begin{alignat*}{3}
  M^c_{11}&= 33.234, & \quad  M^c_{12}&= -353.598,  
  & \quad  M^c_{22}&= 3762.753.
\end{alignat*}
 \item[(1-3)] $\sqrt{s}=10$ TeV
\begin{alignat*}{3}
  M^c_{11}&= 59.333, & \quad  M^c_{12}&=-632.179,  
  & \quad  M^c_{22}&= 6736.735.
\end{alignat*}
 \item[(1-4)] $\sqrt{s}=14$ TeV
\begin{alignat*}{3}
  M^c_{11}&= 134.052, & \quad  M^c_{12}&= -1428.300,  
  & \quad  M^c_{22}&=15220.286.
\end{alignat*}
\end{description}
 \item[(2)] The energy distribution
 \begin{description}
 \item[(2-1)] $\sqrt{s}=7$ TeV
 \begin{alignat*}{3}
  M^E_{11}&= 23.102, & \quad  M^E_{12}&= -245.412,  
  & \quad  M^E_{13}&= 0.000,\\
  M^E_{22}&= 2607.658, & \quad  M^E_{23}&= -1.974,  
  & \quad  M^E_{33}&=15.323. 
\end{alignat*}
 \item[(2-2)] $\sqrt{s}=8$ TeV
  \begin{alignat*}{3}
  M^E_{11}&= 33.234, & \quad  M^E_{12}&= -353.598,  
  & \quad  M^E_{13}&= 0.000,\\
  M^E_{22}&= 3763.252, & \quad  M^E_{23}&= -2.880,  
  & \quad  M^E_{33}&=21.124. 
\end{alignat*}
 \item[(2-3)] $\sqrt{s}=10$ TeV
\begin{alignat*}{3}
  M^E_{11}&=59.333, & \quad  M^E_{12}&= -632.179,  
  & \quad  M^E_{13}&= 0.000,\\
  M^E_{22}&=6737.696, & \quad  M^E_{23}&= -5.120,  
  & \quad  M^E_{33}&=35.233. 
\end{alignat*}
 \item[(2-4)] $\sqrt{s}=14$ TeV
\begin{alignat*}{3}
  M^E_{11}&= 134.052, & \quad  M^E_{12}&= -1428.300,  
  & \quad  M^E_{13}&= 0.000,\\
  M^E_{22}&= 15222.443, & \quad  M^E_{23}&= -10.953,  
  & \quad  M^E_{33}&= 72.264. 
\end{alignat*}
\end{description}
\end{description}
Here, all $M_{13}^E$ became zero for the same reason as vanishing $d_R$ terms
in the angular distribution, i.e., 
the decoupling theorem~\cite{Grzadkowski:1999iq}-\cite{Godbole:2006tq}.
Using the inverse matrices calculated from the above elements 
and Eq.(\ref{eq:uncertainty1}) -- Eq.(\ref{eq:uncertainty3}),
we can estimate the statistical uncertainties of the relevant couplings.

Before giving the results, however, we should comment about a stability problem
in inverse-matrix computations: We noticed that the results fluctuate to a certain
extent (beyond our expectation) depending on to which decimal places of
$M^{c,E}$ we take into account as our input data.
Therefore we calculate $X^{c,E}$ for the above $M^{c,E}$ and also for what
are obtained by rounding those $M^{c,E}$ off to two decimal places.
We then use their mean values to get
$\delta d_V$ and $\delta d_R$ with ``errors'', which are the differences
between the mean values and the maximum/minimum.
Let us explain these treatments more specifically by taking (1-1) as an example:
For those different inputs, we get
\begin{eqnarray*}
&&\sqrt{X_{22}^c} = 1.73\ \ {\rm for}\ \ M_{11}^c=23.102,\,\ \cdots,      \\
&&\phantom{\sqrt{X_{22}^c}}
                  = 2.57\ \ {\rm for}\ \ M_{11}^c=23.10,\phantom{2}\ \cdots, 
\end{eqnarray*}
which lead to the mean value 2.15, the maximum 2.57, and the minimum 1.73, and
resultant errors $\pm 0.42$, which are derived by $2.57-2.15$ and $1.73-2.15$.
These uncertainties are of course independent of the statistical ones which the
OOA gives, i.e., $\delta d_{V,R}$ themselves, so we should show both uncertainties
separately as different errors.

This instability is due to the large contribution from the $d_V$ coupling, which
is dependent of gluon momenta and could therefore dominate. That is, we are consequently
forced to treat large and small figures simultaneously in one matrix. In fact,
we confirmed that any similar problem does not appear if we carry out computations
for $M^E$ and $X^E$ without $d_V$ terms, in which there is no big difference between
the SM- and $d_R$-term contributions.

We also have to explain how we can take into account QCD higher-order corrections: All
the numerical computations of $M_{ij}^{c,E}$ in this section were done with the
tree-level formulas. In order to include QCD corrections there, we multiply the
tree cross sections by the $K$-factor. This factor disappears in the combination
$X_{ii}^{c,E}\sigma_{\ell}$ and remains only in $N_{\ell}\,(=L\sigma_{\ell})$ when
we estimate $\delta d_{V,R}$. Therefore the luminosity $L$ discussed in the following
should be understood as an effective one including $K$ (and also the lepton detection
efficiency $\epsilon_\ell$).

We now show the whole results in Table~\ref{tab:ang} and Table \ref{tab:ene}
for the angular and energy distributions, respectively. 
The uncertainties we encountered here are hard to eliminate, but still the results
will tell us the necessary luminosity for reaching the precision
which we aim to realize.
For example, we need at least $L \simeq 1500~{\rm pb}^{-1}(=1.5 ~{\rm fb}^{-1})$ in order to achieve
$|\delta d_V|\simeq O(10^{-2})$ in case of measuring the angular distribution
at the LHC whose colliding energy is 8 TeV.
As mentioned, those $L$ should be divided by $K \simeq 1.5$ and $\epsilon_\ell$.
If we assume $\epsilon_\ell = 0.5$, the resultant $L$ increases slightly, which
however hardly affects our conclusion.

%
%
\vspace*{0.5cm}
\begin{table}[H]
\begin{minipage}{0.40\hsize}
\begin{center}
\begin{tabular}{ c|c }
$\sqrt{s}$~[TeV] & $|\delta d_V| \!\times\! {\sqrt{L}}  $ \\ \hline
 7&$2.15 \pm 0.42$\\ 
 8&$2.25 \pm 0.96$\\ 
 10&$1.12 \pm 0.12 $\\ 
 14&$0.73 \pm 0.02 $\\ \hline
\end{tabular}
\caption{Estimated statistical uncertainties of $d_V$  (multiplied by $\sqrt{L}$) from
the angular distribution of a decay lepton.}\label{tab:ang}
\end{center}
\end{minipage}
\hspace*{0.5cm}
\begin{minipage}{0.55\hsize}
%
\begin{center}
\begin{tabular}{c|c|c} 
$\sqrt{s}$~[TeV]&$|\delta d_V| \! \times \! {\sqrt{L}} $ &$|\delta d_R| \! \times \! {\sqrt{L}} $ \\ \hline
7 &$1.86\pm 0.28 $ &$0.35 \pm 0.02 $ \\ 
8 &$1.70\pm 0.51 $ &$0.32 \pm 0.05 $ \\ 
10&$0.98\pm 0.08 $ &$0.22 \pm0.01 $ \\
14&$0.65 \pm 0.02$ &$0.15 $ \\\hline
\end{tabular}
\caption{Estimated statistical uncertainties of $d_V$ and $d_R$ (multiplied by $\sqrt{L}$)
 from the energy distribution of a decay lepton.}\label{tab:ene}
\end{center}
\end{minipage}
\end{table}

\vskip 0.5cm
Here we performed all the angular and energy integrations to compute $M^{c,E}$ over the full
kinematically-allowed ranges for simplicity, but it will be useful to see how our results are
affected by specific experimental conditions. As a typical example, we take the transverse-momentum
cut for $\ell^+$ in the energy distribution: $p_{\ell\,{\rm T}} \ge p_{\ell\,{\rm T}}^{min}$,
which leads to
\[
-\sqrt{1-(p_{\ell\,{\rm T}}^{min}/E_{\ell})^2} \le \cos\theta_{\ell}
\le +\sqrt{1-(p_{\ell\,{\rm T}}^{min}/E_{\ell})^2}
\]
with $E_{\ell} \ge p_{\ell\,{\rm T}}^{min}$.
The results for $p_{\ell\,{\rm T}}^{min}=$20 GeV (see, e.g., \cite{Aad:2012qf,Aaltonen:2011zma})
are
\begin{equation}
|\delta d_V|\,(\times {\sqrt{L}}) = 2.30 \pm 0.51
\ \ \ {\rm and}\ \ \ 
|\delta d_R|\,(\times {\sqrt{L}}) = 0.69 \pm 0.12
\end{equation}
for $\sqrt{s}=$8 TeV, where our results with no experimental cuts were most unstable.
We find there is no change in the instability of $\delta d_V$ while that of $\delta d_R$ has
increased. However, the relative size of it compared to $\delta d_R$ itself (= 0.69) is almost
the same as that in the Table \ref{tab:ene}: 16 \% $\to$ 17 \% (30 \% $\to$ 22 \% in $\delta d_V$).
Therefore, they tell us that introducing a lepton-$p_{\rm T}$ cut will not seriously affect the
stability of $X^{c,E}$, although the expected uncertainties themselves become a bit larger, which
we can take into account through the detection efficiency as will be discussed in the last section.

%
\sec{Summary}
%
Possible non-standard top-gluon and top-$W$ couplings were studied for hadron-collider
experiments in a model-independent way.
Those couplings are derived as parameters which characterize the effects
of dimension-6 effective operators based on a scenario of Buchm\"{u}ller
and Wyler~\cite{Buchmuller:1985jz}.
More specifically, we studied $d_{V,A}$ couplings which contribute to top-pair productions 
and $d_R$ coupling which affects leptonic decays of a top quark as non-standard interactions.

First, we updated the bounds of $d_V$ and $d_A$ from the current Tevatron
and LHC data of top-pair productions and showed that the allowed area of 
$d_{V,A}$ became narrower around ($d_V$,$d_A$)=(0,0).
Based on this observation, we adopted a linear approximation for the non-standard
parameters in the subsequent optimal-observable analysis (OOA) of $pp \to t\bar{t}X \to \ell^+ X'$. 
In this approximation, all the contributions from $d_A$ were neglected 
because the leading order of $d_A$ terms is quadratic.  
Therefore, the OOA was performed for $d_V$ and $d_R$ 
to estimate their expected statistical uncertainties
using the angular and energy distributions of the final lepton.

Measuring the energy distribution could study both $d_V$ and $d_R$ at the 
same time with a higher precision than the case of the angular distribution.
Meanwhile, since the angular distribution is affected by only $d_V$ contributions
within our approximation,
this one is suitable for exploring the mechanism of top-pair productions exclusively though
the precision of $d_V$ is slightly lower than that from the energy distribution.

We found that there were some ambiguities in inverse-matrix computations depending on
where to round the input data (i.e., $M_{ij}^{c,E}$-elements) off, and consequently
estimated statistical errors were affected thereby to a certain extent.
However, our conclusion about the necessary luminosity and new-physics scale
${\mit\Lambda}$ hardly receive that serious influence, because those ambiguities
do not change the order of the results.

As for future prospects, if the LHC is going to be steadily upgraded and
$L=500 ~{\rm fb}^{-1}$ is achieved at $\sqrt{s}=14$ TeV,
both $d_V$ and $d_R$ could be determined with
$|\delta d_V| \sim O(10^{-3})$  and $|\delta d_R| \sim O(10^{-4})$
from the energy distribution:
Here, 
$|\delta d_V| \sim 0.001$ $(|\delta d_R| \sim 0.0001)$ means that
contribution from effective operator ${\cal O}_{uG\phi}^{33}~ ({\cal O}_{uW}^{33})$
is suppressed by ${\mit\Lambda} \gtrsim 7 ~(20)$ TeV 
if the center value of measured $d_V$ ($d_R$) is close to zero 
and $C^{33}_{uG\phi} ~(C_{uW}^{33}) \sim 1$, which we estimated from 
Eqs.(\ref{eq:dvdadef}) and (\ref{eq:drdef}).
%

Finally, we did not discuss here any realistic experimental conditions
like effects of event-selection cuts and/or backgrounds except for
the discussion in the last paragraph of the preceding section. This was partly
because it was not our purpose to go into such details in this letter,
although it is important since they decrease actual event number,
but also because the OOA can work under any experimental conditions.
All we have to do is to take account of them properly and accept
the estimated luminosity as an effective one including all those effects.
This is closely related to the detection efficiency $\epsilon_\ell$,
which we discussed at the end of the previous section. Indeed we are ready
to re-calculate every quantity obtained in this letter, correspondingly to
any lepton angle- and/or energy-cut conditions since we gave
all the formulas used here analytically in Ref.\cite{HIOKI:2011xx}.

%
\secnon{Acknowledgments}
%
This work was partly supported by the Grant-in-Aid for Scientific Research 
No. 22540284 from the Japan Society for the Promotion of Science.
Part of the algebraic and numerical calculations were carried 
out on the computer system at Yukawa Institute for Theoretical
Physics (YITP), Kyoto University.
\baselineskip=20pt plus 0.1pt minus 0.1pt

 \vspace*{0.8cm}

\end{document}